\documentclass[12pt]{article}
\usepackage{epsfig}
\usepackage{latexsym}
\DeclareGraphicsExtensions{.eps.gz,.eps,.ps,.ps.gz}
\parskip=\itemsep               
\newcommand{\AmS}{{\protect\the\textfont2
  A\kern-.1667em\lower.5ex\hbox{M}\kern-.125emS}}
\newcommand{\lwig}{\mbox{\,\raisebox{.3ex}
    {$<$}$\!\!\!\!\!$\raisebox{-.9ex}{$\sim$}\,}}

\def\Journal#1#2#3#4{{#1} {#2} (#4) #3}

\def\NPB{Nucl. Phys. B}
 
\def\PLB{Phys. Lett. B} 
\def\PRL{Phys. Rev. Lett.} 
\def\PRD{Phys. Rev. D}

\date{\empty}
\title{
{\normalsize\rightline{DESY 99-062}\rightline{hep-ph/9905384}} 
\vskip 1cm 
      \bf Instantons in the QCD Vacuum and\\
       \bf in Deep Inelastic Scattering\thanks{Talk presented at the 
    7th International Workshop on Deep Inelastic Scattering and QCD (DIS\,99),
    Zeuthen/Germany, April 19-23, 1999; to be published in the Proceedings 
    (Nuclear Physics B (Proc. Suppl.)).} 
       \vspace{21mm}}
\author{A. Ringwald and F. Schrempp\\[4mm] 
Deutsches Elektronen-Synchrotron DESY, Hamburg, Germany}
\begin{document}
\begin{titlepage} 
  \maketitle
\begin{abstract}
We give a brief status report on our on-going investigation of the
prospects to discover QCD instantons in deep inelastic scattering (DIS) 
at HERA. A recent high-quality lattice study of the
topological structure of the QCD vacuum is exploited to provide
crucial support of our predictions for DIS, based on instanton
perturbation theory.  
\end{abstract}


\thispagestyle{empty}
\end{titlepage}
\newpage \setcounter{page}{2}

\section{INTRODUCTION}
The ground state (``vacuum'') of non-abelian gauge theories like QCD
is known to be very rich. It includes {\it topologically non-trivial}
fluctuations of the  gauge fields, carrying an integer topological
charge Q. The simplest building blocks of topological structure in the 
vacuum, localized (i.\,e. ``instantaneous'') in (euclidean) time  and   
space are~\cite{bpst,th} {\it instantons} ($I$\,) with $Q=+1$ and {\it
anti-instantons} ($\overline{I}$\,) with $Q=-1$. 
While they are believed to play an important r{\^o}le in various
long-distance aspects of QCD, there are also important {\it
short-distance} implications. In QCD with $n_f$ (massless) flavours,
instantons  induce {\it hard} processes violating
``chirality'' $Q_5$ by an amount $\Delta Q_5=2\,n_f\,Q$, in accord
with the general ABJ chiral 
anomaly relation~\cite{th}. While in ordinary perturbative QCD ($Q=0$), these
processes are forbidden, their experimental discovery would clearly be of basic
significance.   The DIS regime is strongly 
favoured in this respect, since {\it hard} $I$-induced processes are both
calculable~\cite{mrs1,rs-pl} within $I$-pertubation theory and have good
prospects for experimental detection at HERA~\cite{rs-pl,rs,grs,cgrs}.
\section{INSTANTONS IN THE QCD VACUUM}      
\begin{figure}
\parbox{6cm}{\epsfig{file=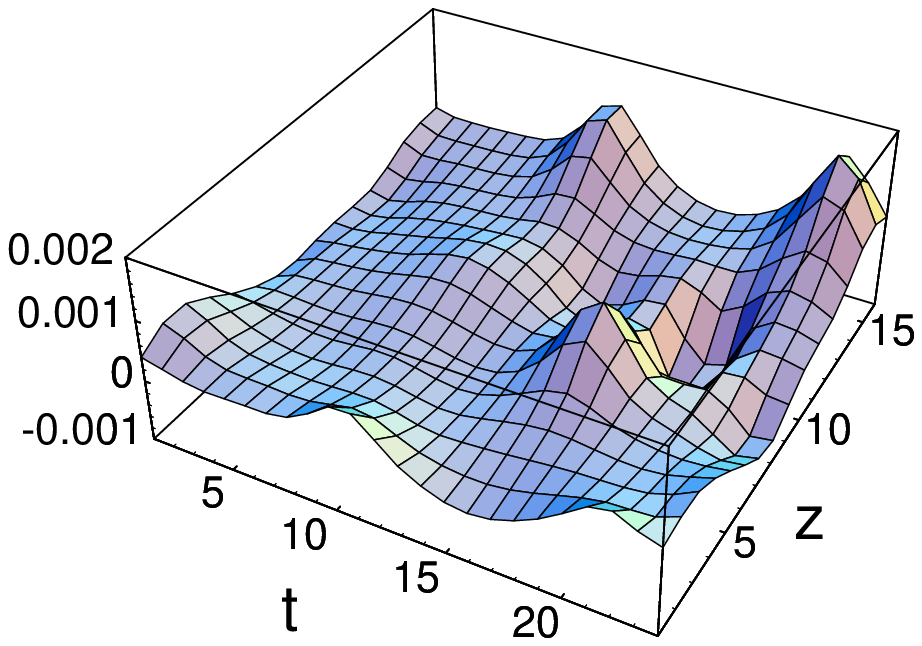,width=6cm}}
\parbox{5cm}{\vspace{-3ex}\epsfig{file=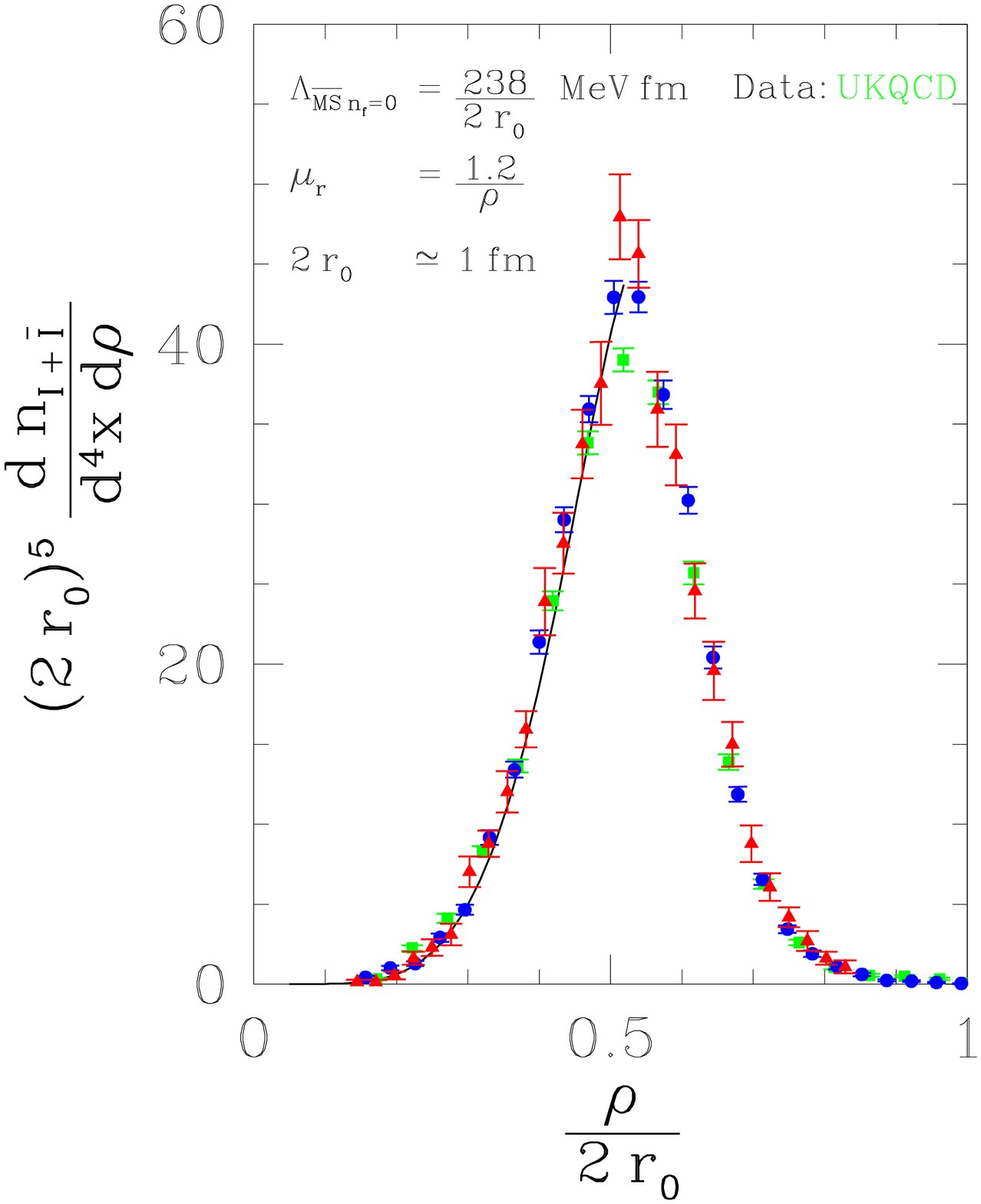,width=5cm}}
\parbox{4.5cm}{\epsfig{file=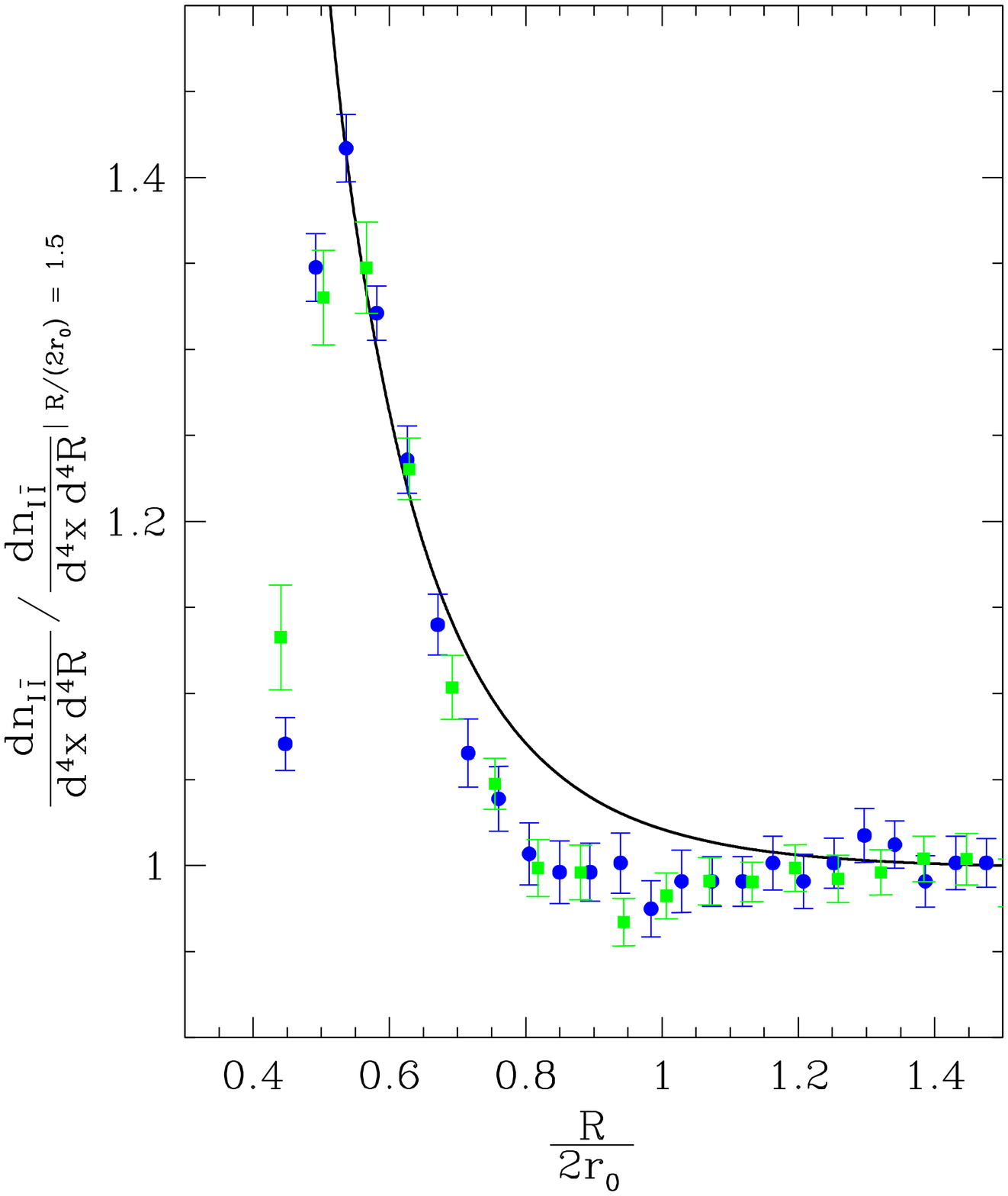,width=4.5cm}}
\caption{(left): Topological charge density $q(\vec{x},t)$ on the
lattice~\cite{chu} after ``cooling'', displayed as function of $z$ and
$t$ with $x,y$  fixed. Three $I$'s  ($q(\vec{x},t)>0$)  and two
$\overline{I}$'s ($q(\vec{x},t)<0$) are visible as bumps.   
Continuum limit~\cite{rs-lat} of ``equivalent'' UKQCD
data~\cite{ukqcd,mike} for the ($I+\overline{I}$)-size distribution (middle)  
and the normalized $I\overline{I}$-distance distribution (right) along with the
respective predictions from $I$-perturbation theory and the valley
form of the $I\overline{I}$-interaction~\cite{rs-lat}. The 3-loop form of
$\alpha_{\rm \overline{MS}}$ with $\Lambda_{{\rm
\overline{MS}}\,n_f=0}$ from ALPHA~\cite{alpha} was used.  
} 
\label{figone}
\end{figure}

Crucial information~\cite{rs-lat} on the range of validity of our DIS
predictions~\cite{mrs1,rs-pl} 
comes from a recent high-quality lattice investigation~\cite{ukqcd} on the
topological structure of the QCD vacuum (for $n_f=0$).
In order to make $I$-effects visible in lattice simulations with
given lattice spacing $a$, the raw data have to be ``cooled'' first.
This procedure is to filter out
(dominating) fluctuations of {\it short} wavelength ${\cal O}(a)$,
while affecting the topological fluctuations of much longer wavelength 
$\rho \gg a$ comparatively little. After cooling, an ensemble of $I$'s and
$\overline{I}$'s can clearly be seen (and studied) as bumps in the
topological charge density (e.g. fig.~\ref{figone}\,(left)) and 
in the Lagrange density.  

Next, we note that crucial $I$-observables in DIS, like the
$I$-induced rate at HERA, are closely related to $I$-observables in
the QCD vacuum, as measured in lattice simulations. 

The link is provided through two basic quantities of the $I$-calculus,
$D(\rho)$, the $I$-size distribution and
$\Omega(U,R^2/\rho\overline{\rho},\overline{\rho}/\rho)$, the
$I\overline{I}$-interaction. Here $\rho (\overline{\rho}), R_\mu$ and
the matrix $U$ denote the $I\,(\overline{I})$-sizes, the
$I\overline{I}$-distance 4-vector and the $I\overline{I}$ relative
color orientation, respectively. Within $I$-perturbation theory,
the functional form of $D$ and $\Omega$ is {\it known} for
$\alpha(\mu_r)\log(\mu_r\rho)\ll 1$ and $R^2/\rho\overline{\rho}\gg
1$, respectively, with $\mu_r$ being the renormalization scale. Within 
the so-called ``$I\overline{I}$-valley''
approximation~\cite{yung,valley-most-attr-orient,valley-gen-orient},
$\Omega_{\rm valley}$ is even analytically known for all $R^2$. 

Fig.~\ref{figone}\,(middle) illustrates the striking agreement in shape and
normalization~\cite{rs-lat} of $2\,D(\rho)$ with the continuum limit of the
high-quality UKQCD lattice data~\cite{ukqcd} for
$d n_{I+\overline{I}}/d^4x\,d\rho$.   
The predicted normalization of $D(\rho)$ is very sensitive to $\Lambda_{{\rm
\overline{MS}}\,n_f=0}$ for which we took the most accurate
(non-perturbative) result from ALPHA~\cite{alpha}. The theoretically favoured
choice $\mu_r\rho={\cal O}(1)$ in fig.~\ref{figone}\,(middle),
optimizes the range of agreement, extending right up to the peak
around $\rho\simeq 0.5$ fm. However, due to its two-loop
renormalization-group invariance, $D(\rho)$ is almost independent
of $\mu_r$ for $\rho \lwig 0.3$ fm over the large range
$2\lwig\mu_r\lwig 20$ GeV. Hence  for $\rho \lwig 0.3$ fm, there is
effectively no free parameter involved! 

Fig.~\ref{figone}\,(right) displays the continuum
limit~\cite{rs-lat} of the UKQCD data~\cite{ukqcd,mike} for the
 distance distribution of $I\overline{I}$-pairs,
$dn_{I\overline{I}}/d^4x\,d^4R$,  
along with the theoretical prediction~\cite{rs-lat}. The latter
involves (numerical) integrations of $\exp
(-4\pi/\alpha\cdot\Omega_{\rm valley})$ over the
$I\overline{I}$ relative color orientation $(U)$, as well as $\rho$ and
$\overline{\rho}$. For the respective weight $D(\rho)D(\overline{\rho})$,  
a Gaussian fit to the lattice data was used in order to avoid convergence
problems at large $\rho,\overline{\rho}$. We note a good
agreement with the lattice data down to $I\overline{I}$-distances
$R/\langle\rho\rangle\simeq 1$. These results imply first direct
support for the validity of the ``valley''-form of the interaction
$\Omega$ between $I\overline{I}$-pairs. 
 
In summary: The striking agreement of the UKQCD lattice data with
$I$-perturbation theory is a very interesting result by itself. The
extracted lattice constraints on the range of validity of
$I$-perturbation theory can be directly translated into a ``fiducial''
kinematical region for our
DIS-predictions~\cite{rs-pl,rs-lat}. Our results also suggest a 
promising proposal~\cite{rs-lat}: One may try and replace the  
two crucial quantities of the perturbative $I$-calculus
$D(\rho)$  and $\Omega(U,R^2/\rho\overline{\rho},\overline{\rho}/\rho)$
by their actual form inferred from the lattice data. The present
``fiducial'' cuts in DIS may
then be considerably relaxed, high-$E_T$ photoproduction becomes
accessible theoretically, etc. 

\section{SEARCH STRATEGIES IN DIS}



An indispensable tool for investigating the prospects to detect
$I$-induced processes at HERA, is our
$I$-event generator~\cite{grs} QCDINS-1.60, which 
is interfaced (by default) to HERWIG 5.9.   

In a recent detailed study~\cite{cgrs}, based on QCDINS and standard DIS 
event generators, a number of basic (experimental) questions has been
investigated:  How to  isolate an {\it
$I$-enriched} data sample by means of cuts to a set of observables?
How large are the dependencies on Monte-Carlo models, both for $I$-induced
(INS) and normal DIS events? Can the Bjorken-variables 
$(Q^\prime,\ x^\prime)$ of the $I$-subprocess(, to which 
``fiducial'' cuts should be applied,) be reconstructed? 

\begin{figure} [h]
\begin{center}
\epsfig{file=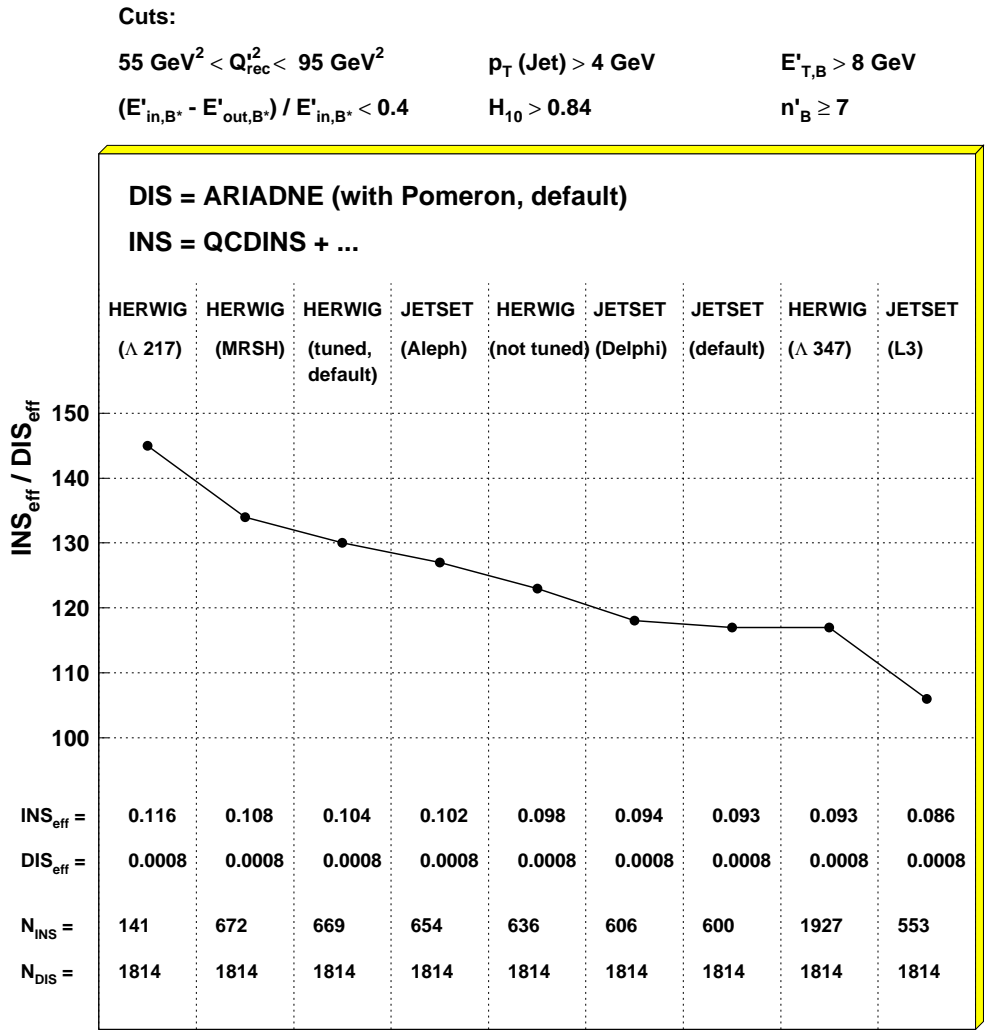,width=13cm}
\end{center}
\caption{Dependence of the $I$-separation power for a multi-dimensional
cut scenario on the variation of Monte-Carlo models and
parameters~\cite{cgrs}. Corresponding efficiencies (eff) and event
numbers for $\int {\cal L} dt =30\ {\rm pb}^{-1}$ are also shown. DIS
= default, INS varied.}
\label{figtwo}
\end{figure}

\begin{figure}
\begin{center} 
\epsfig{file=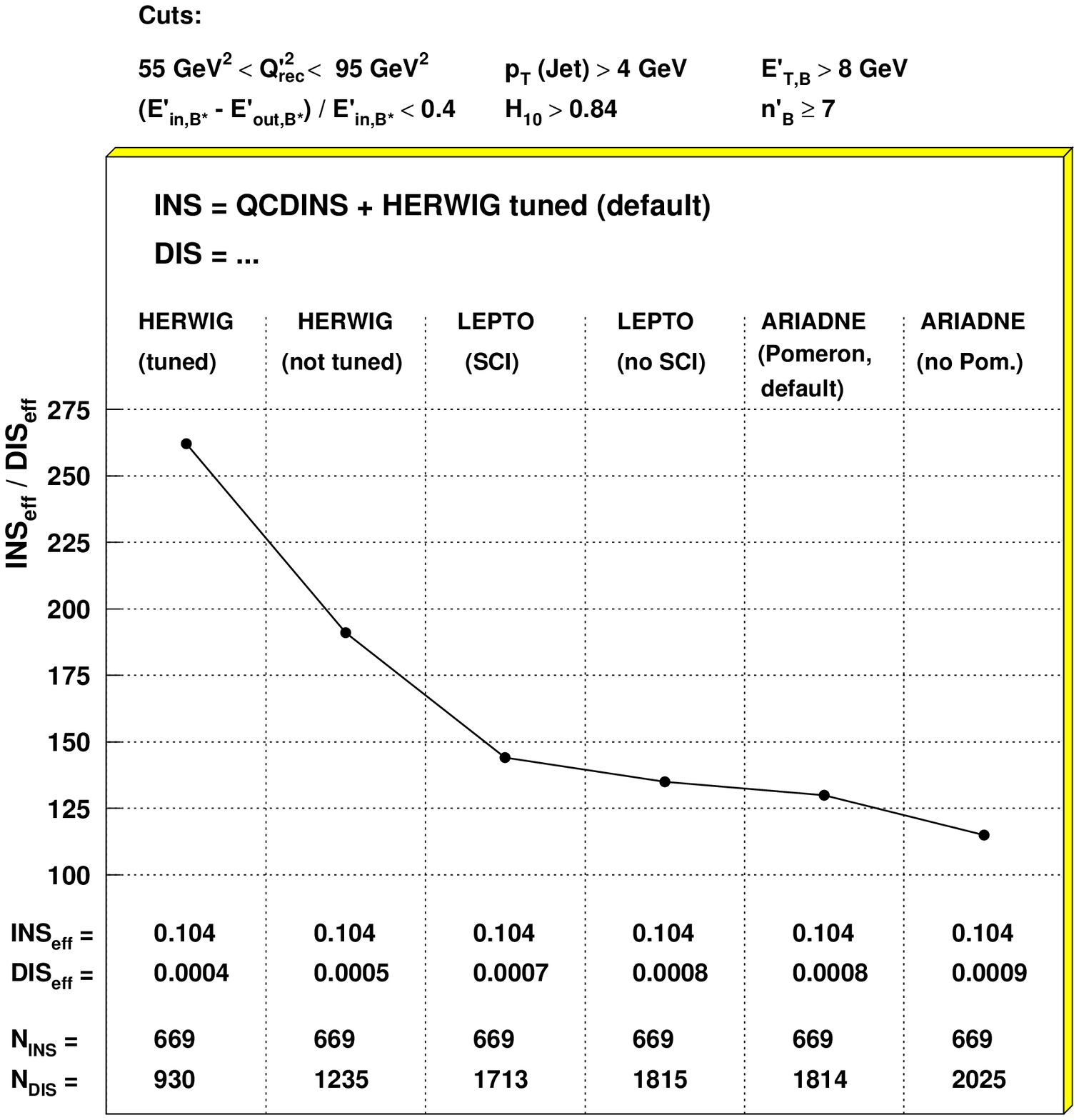,width=13cm}
\end{center}
\caption{As in fig.~\ref{figtwo}, but INS = default, DIS varied.}
\label{figthree}
\end{figure}

Let us briefly summarize the main results. 
While the ``$I$-separation power''= $\rm INS_{\rm
eff}/DIS_{\rm eff}$ typically ranges around ${\cal O}(20)$ for {\it single}
observables, a set of six observables (among $\sim 30$) with much improved
$I$-separation 
power $={\cal O}(130)$  could be found.  The systematics
induced by varying the modelling of $I$-induced events remains
surprisingly small (fig.~\ref{figtwo}). In contrast,   
the modelling of normal DIS events in the relevant region of phase
space turns out to depend quite strongly on the used generators and
parameters (fig.~\ref{figthree}). Despite a relatively high
expected rate of ${\cal O}(100)$ pb for $I$-events in the ``fiducial''
DIS region~\cite{rs-pl}, a better understanding of the tails of
distributions for normal DIS events turns out to be quite important. 


\begin{thebibliography}{9}
\bibitem{bpst}
A. Belavin, A. Polyakov, A. Schwarz and Yu. Tyupkin, 
\Journal{\PLB}{59}{85}{1975}.
\bibitem{th} 
G. `t Hooft, \Journal{\PRL}{37}{8}{1976};
\Journal{\PRD}{14}{3432}{1976}; \Journal{\PRD}{18}{2199}{1978}
(Erratum). 
\bibitem{mrs1}
S. Moch, A. Ringwald and F. Schrempp, \Journal{\NPB} {507}{134}{1997}.
\bibitem{rs-pl}
A. Ringwald and F. Schrempp, \Journal{\PLB} {438}{217}{1998}.
\bibitem{rs}
A. Ringwald and F. Schrempp, hep-ph/\-9411217, in:
{\it Quarks `94}, Proc. 8th Int. Seminar, Vladimir, Russia, 
1994, pp. 170-193.
\bibitem{grs}
M. Gibbs, A. Ringwald and F. Schrempp, hep-ph/9506392,
in: {\it Proc. DIS95}, Paris, 1995, pp. 341-344.
\bibitem{cgrs}
T. Carli, J. Gerigk, A. Ringwald and F. Schrempp, to appear.
\bibitem{rs-lat}
A. Ringwald and F. Schrempp, hep-lat/\-9903039.
\bibitem{ukqcd}
D.A. Smith and M.J. Teper, (UKQCD), \Journal{\PRD}{58}{014505}{1998}. 
\bibitem{chu}
M.-C. Chu, J.M. Grandy, S. Huang and J.W. Negele,
\Journal{\PRD}{49}{6039}{1994}.
\bibitem{yung}
A. Yung,
\Journal{\NPB}{297}{47}{1988}.
\bibitem{valley-most-attr-orient}
V.V. Khoze and A. Ringwald, 
\Journal{\PLB}{259}{106}{1991}.
\bibitem{valley-gen-orient}
J. Verbaarschot, 
\Journal{\NPB}{362}{33}{1991}.
\bibitem{mike}
 M. Teper, private communication.
\bibitem{alpha}
S. Capitani, M. L{\"u}scher, R. Sommer and H. Wittig,
\Journal{\NPB}{544}{669}{1999}.
\end{thebibliography}
\end{document}